\documentclass[12pt,a4paper]{article}
\usepackage{amstext,amssymb,amsmath,amscd,amsfonts}

\input{tcilatex}

\begin{document}

\title{Non-Classical Expected Utility Theory\thanks{%
The financial support of the grant \#NSh-6417.2006.6, School Support, is
gratefully acknowledged. }}
\author{ V. I. Danilov\thanks{%
Central Economic Mathematical Institute, Russian Academy of\ Sciences, 47
Nakhimovski Prospect, 117418 Moscow, Russia. danilov@cemi.rssi.ru} \ and A.
Lambert-Mogiliansky\thanks{%
PSE,\ Paris-Jourdan Sciences Economiques (CNRS, EHESS, ENS, ENPC),\ Paris,\
alambert@pse.ens.fr}}
\date{}
\maketitle

\begin{abstract}
In this paper we extend Savage's theory of decision-making under uncertainty
from a classical environment into a non-classical one. We formulate the
corresponding axioms and provide representation theorems for qualitative
measures and expected utility.
\end{abstract}

\section{Introduction}

In this paper we propose an extension of the standard approach to
decision-making under uncertainty in Savage's style from the classical model
into the more general model of non-classical measurement theory
corresponding the basic mathematical structure of Quantum Mechanics (see 
\cite{DLM}). .

Formally, this means that we substitute the Boolean algebra model with a
more general ortholattice structure (see \cite{DLM}). In order to provide a
first line of motivation for our approach we turn back to Savage's theory in
a very simplified version. In Savage \cite{Sav}, the issue is about the
valuation of ``acts'' with uncertain consequences or results. One of our
simplifications is to assume that we are able to provide a cardinal measure
of the results - in utils (we below clarify this somehow imprecise
formulation). Acts lead to results (measurable in utils), but the results
are uncertain. How can one formalize acts with uncertain outcomes?

The classical approach amounts to the following. There exists a set $X$ of
states of nature, which may in principle occur. An act corresponds to a
function $f:X\to \mathbb{R}$. If the state $s\in X$ is realized, our agent
receives a utility of $f(s)$ utils. But before hand it is not possible to
say which state $s$ is going to be realized. To put it differently, the
agent has to choose among acts \textit{before} he learns about the state $s$%
. This is the heart of the problem.

Among possible acts there are ``constant'' acts, i.e., acts with a result
that is known before hand, independently of the state of nature $s$. The
constant act is described by a (real) number $c\in \mathbb{R}$. It is
therefore natural to link an arbitrary act $f$ with its ``utility
equivalent'' $CE( f) \in \mathbb{R}$. This corresponds to defining the
constant act $c\ $(with utility outcome $CE( f) $) such that our
decision-maker is indifferent between the act $f$ and the constant act (with
utility value) $CE(f)$. The first postulate of this simplified (we assume
also that the set $X$ is finite) Savage model asserts the existence of the 
\textit{certainty equivalent}:

\begin{itemize}
\item \textit{S1.} There exists a certainty equivalent $CE:\mathbb{R} ^{X}
\to \mathbb{R}$ and for the constant act $1_X$ we have $CE(1_X)=1$.
\end{itemize}

\noindent It is rather natural to require monotonicity of the mapping $CE$:

\begin{itemize}
\item \textit{S2.} If $f\leq g$ then $CE( f) \leq CE( g) $.
\end{itemize}

\noindent The main property we impose on $CE$ is linearity:

\begin{itemize}
\item \textit{S3.} $CE( f+g) =CE( f) +CE( g) $ for any $f$ and $g\in \mathbb{%
R}^{X}.$
\end{itemize}

\medskip

Presented in such a way, this requirement looks like a very strong condition
indeed. Savage himself and his followers preferred to "hide" it behind the
so-called "sure thing principle" so that the linearity is derived from some
other axioms. But as we understand it, this is more of an artifice.

In fact axiom $S3$ should be understood as a condition of \textit{additivity}
rather linearity. But together with monotonicity axiom $S3$ implies true
linearity, that is $CE(\alpha f+\beta g)=\alpha CE(f)+\beta CE(g)$ for any $%
\alpha ,\beta \in \mathbb{R}$. As a linear functional on the vector space $%
\mathbb{R}^{X}$, $CE$ can be written in a form $CE(f)=\sum_{x}f(x)\mu (x)$.
By axiom $S2$, $\mu \geq 0$; by $CE(1_{X})=1$ we have $\sum_{x}\mu (x)=1$.
Therefore $\mu (x)$ can be interpreted as the \textquotedblleft
probability\textquotedblright\ for the realization of state $x$. Sometimes
this probability is called subjective or personal, because it only expresses
the likelihood that a specific decision-maker assigns to event $x$. With
such an interpretation, $CE(f)$ becomes the "expected" utility of the
uncertain act $f$ (or it is better to say - of the act $f$ with uncertain
outcome).

With such a view we may assign probabilities not only to single state $x$
but also to any subset of states (or to any event) $A\subset X$. $\mu (A)$
can be understood either as the sum $\sum_{x\in A}\mu (x)$, or as $CE(1_{A})$%
, where $1_{A}$ is the characteristic function of subset $A$. The
interpretation in the second approach is clear: the act $1_{A}$ is a bet on
event $A$ such that we receive $1$ util if event $A$ is realized and $0$
util otherwise (if the opposite or complementary event $\overline{A}$
occurs). The decision-maker can compare such bets on events and thereby
compare events with respect to their likelihood. So we arrive at the notion
of qualitative probability measure.

Our main idea is to substitute the Boolean lattice of events with a more
general ortholattice. The move in that direction was initiated long ago, in
fact with the creation of Quantum Mechanics. The Hilbert space entered into
the theory immediately, beginning with von Neumann \cite{vN} who proposes
the lattice of projectors in the Hilbert space as a suitable model instead
of the classical (Boolean) logic. Birkhoff and von Neumann in their seminal
paper \cite{BvN} have investigated the necessary properties of such a
non-distributive logic (modularity, ortho-modularity?). Recently a few
decision-theoretical papers appear (see for example, \cite{Deut, Pit, GH,
LS, LaMura}) in which the standard expected utility theory was transposed
into Hilbert space model. Beside the formal arguments, a motivation for this
research is that a more general description of the world allows to explain
some behavioral anomalies e.g., the Eldsberg paradox (see \cite{GH}).

Lehrer and Shmaya write \textquotedblleft We adopt a similar approach and
apply it to the quantum framework... While classical probability is defined
over subsets (events) of a state space, quantum probability is defined over
subspaces of a Hilbert space." Gyntelberg and Hansen (2004) apply a general
event-lattice theory (with axioms that resemble those of von Neumann and
Morgenstern) to a similar framework. One could expect that Gyntelberg and
Hansen truly would have been working with general ortholattices. But no,
they also worked with subspaces of a Hilbert space. Our first aim is to show
that there is no need for a Hilbert space, the Savage approach can just as
well (and even easier) be developed within the frame of more general
ortholattices. Another line of motivation that we share with other decision
theoretical papers in this vein is that this model maybe a better
representation of the subjectively perceived world. For non-classical
features of perception see for instance \cite{atman}.

\section{Ortholattices}

A \emph{lattice} is an ordered set such that any of its subsets (including
the empty subset) has a greatest lower bound ($\vee $ or $\sup $) and a
lowest higher bound ($\wedge $ or $\inf $), which guarantees the existence
of a maximal element $\mathbf{1}$ and a minimal element $\mathbf{0}$%
\footnote{%
It is more natural to call what we just defined, a complete ortholattice.
Usually one only requires the existence of finite bounds. However we shall
not interest us much for the general case, assuming finiteness of $\mathcal{L%
}$.}. An \emph{ortholattice} is a lattice $\mathcal{L}$ equipped with an
operation of \emph{ortho-complementation} $\perp :\mathcal{L}\rightarrow 
\mathcal{L}$. This operation is assumed to be involutive ($a^{\perp \perp
}=a $), to reverse the order ($a\leq b$ if and only if $b^{\perp }\leq
a^{\perp } $) and to satisfy the following property $a\vee a^{\perp }=%
\mathbf{1}$ (or, equivalently, $a\wedge a^{\perp }=\mathbf{0}$).\medskip

\textbf{Example 1.} Let $X$ be a set and $\mathcal{L=}$ $2^{X}$ the set of
all subsets of $X.\ $The order is defined by set-theoretical inclusion. For $%
A\subset X,\ A^{\perp }=X-A,$ is the set-theoretical complement. It is the
classical situation.\medskip

\textbf{Example 2.} Take some finite dimensional Hilbert space $\mathcal{H}$
(over the field of real or complex numbers). Let $\mathcal{L}$\ be the
lattice of vector subspaces of $\mathcal{H}$ and $\perp $ be the usual
orthogonal complementation.

This example is standard in Quantum Mechanics as well as in the above
mentioned works. But it was early understood that the lattice $\mathcal{L}( 
\mathcal{H}) $ (sometimes it is called the lattice of projectors) is endowed
with a number of special properties. We discuss a significantly more general
case in the next example.\medskip

\textbf{Example 3.} Let $( X,\perp ) $ be an \emph{orthospace} that is a set 
$X$ equipped with an irreflexive and symmetric binary relation
(orthogonality) $\perp $. For $A\subset X$ 
\begin{equation*}
A^{\perp }=\{ x\in X,x\perp a\text{ for all }a\in A\}.
\end{equation*}
The sets of the form $A^{\perp }$ are called \textit{orthoclosed} subsets or 
\emph{flats}. When equipped with the relation $\subset $ and the operation $%
\perp $, flats form an ortholattice $\mathcal{F}(X, \perp)$ (for details see 
\cite{DLM}). Moreover almost any (at least any finite) ortholattice has the
form of $\mathcal{F}( X,\perp)$ for a suitable orthospace $( X,\perp)$.

In order to get better acquainted with this subject, let us consider a few
concrete examples.\medskip

a) Assume that all distinct points of $X$ are pairwise orthogonal. Then any
subset of $X$ is orthoclosed and the ortho-complementation coincides with
the usual set-theoretical complementation. That is we obtain the Boolean
model of Example 1.\medskip

b) Let $X$ be consist of four points $r,l,r^{\prime },l^{\prime }$. The
orthogonality relation is represented by the graph below, where we connect
points with a plain line when they are NON-ORTHOGONAL (so that orthogonal
points as "far from each other" remain unconnected).\medskip

\begin{picture}(76.00,25)(-40,0)
\unitlength=.8mm \special{em:linewidth .4pt} \linethickness{0.4pt}
\put(30.00,5.00){\circle{2.00}} \put(45.00,5.00){\circle{2.00}}
\put(60.00,5.00){\circle{2.00}} \put(75.00,5.00){\circle{2.00}}

 \put(32,5){\line(1,0){11}}
\put(47,5){\line(1,0){11}} \put(62,5){\line(1,0){11}}
\put(30.00,8.00){\makebox(0,0)[cc]{$r$}}
\put(45.00,8.00){\makebox(0,0)[cc]{$l'$}}
\put(60.00,8.00){\makebox(0,0)[cc]{$r'$}}
\put(75.00,8.00){\makebox(0,0)[cc]{$l$}}
\end{picture}

The point $r$ is orthoclosed since $r=\{ r^{\prime },l\} ^{\perp }$;
similarly the point $l$ is orthoclosed. There are two other (nontrivial)
flats: the set $\{ r^{\prime },l\} =r^{\perp }$ and $\{ l^{\prime },r\}
=l^{\perp }$. The corresponding ortholattice is represented below

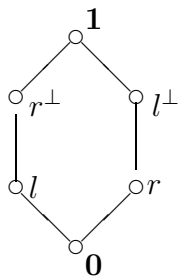
\unitlength=.8mm \special{em:linewidth 0.4pt} \linethickness{0.4pt} 
\begin{picture}(73.00,47.00)(-10,0)
\put(60.00,5.00){\circle{2.00}} \put(70.00,15.00){\circle{2.00}}
\put(50.00,15.00){\circle{2.00}} \put(50.00,30.00){\circle{2.00}}
\put(70.00,30.00){\circle{2.00}} \put(60.00,40.00){\circle{2.00}}

 \put(61,39){\line(1,-1){8}}
 \put(70,29){\line(0,-1){11}}
 \put(69,14){\line(-1,-1){8}}
 \put(51,14){\line(1,-1){8}}
 \put(50,16){\line(0,1){11}}
 \put(51,31){\line(1,1){8}}

\put(63.00,2.00){\makebox(0,0)[cc]{\bf 0}}
\put(63.00,43.00){\makebox(0,0)[cc]{\bf 1}}
\put(53.00,15.00){\makebox(0,0)[cc]{$l$}}
\put(73.00,15.00){\makebox(0,0)[cc]{$r$}}
\put(55.00,29.00){\makebox(0,0)[cc]{$r^\bot$}}
\put(75.00,29.00){\makebox(0,0)[cc]{$l^\bot$}}
\end{picture}\medskip

c) Let us consider the orthospace represented by the following graph

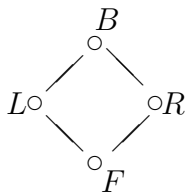
\unitlength=.800mm \special{em:linewidth 0.4pt} \linethickness{0.4pt} 
\begin{picture}(70.00,34.00)(-10,10)
\put(73.00,25.00){\makebox(0,0)[cc]{$R$}}
\put(47.00,25.00){\makebox(0,0)[cc]{$L$}}
\put(62.00,39.00){\makebox(0,0)[cc]{$B$}}
\put(63.00,12.00){\makebox(0,0)[cc]{$F$}}

 \put(52,27){\line(1,1){6}} \put(62,34){\line(1,-1){6}}
\put(69,23){\line(-1,-1){6}} \put(52,23){\line(1,-1){6}}

\put(70.00,25.00){\circle{2.00}} \put(50.00,25.00){\circle{2.00}}
\put(60.00,35.00){\circle{2.00}} \put(60.00,15.00){\circle{2.00}}
\end{picture}

\noindent The corresponding ortholattice is

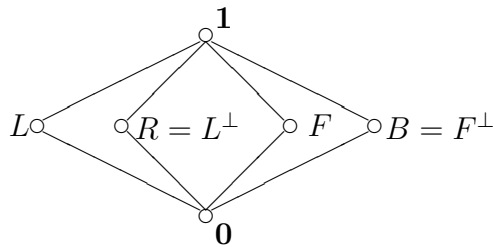
\unitlength=.8mm \special{em:linewidth 0.4pt} \linethickness{0.4pt} 
\begin{picture}(84.00,45.00)(-10,0)
\put(60.00,5.00){\circle{2.00}} \put(46.00,20.00){\circle{2.00}}
\put(32.00,20.00){\circle{2.00}} \put(74.00,20.00){\circle{2.00}}
\put(88.00,20.00){\circle{2.00}} \put(60.00,35.00){\circle{2.00}}

 \put(60,34){\line(-1,-1){13}} \put(60,34){\line(1,-1){13}}
\put(59,34){\line(-2,-1){26}} \put(61,34){\line(2,-1){26}}
\put(59,6){\line(-2,1){26}} \put(60,6){\line(-1,1){13}}
\put(60,6){\line(1,1){13}} \put(61,6){\line(2,1){26}}

\put(63.00,2.00){\makebox(0,0)[cc]{\bf 0}}
\put(63.00,38.00){\makebox(0,0)[cc]{\bf 1}}
\put(29.00,20.00){\makebox(0,0)[cc]{$L$}}
\put(57.00,20.00){\makebox(0,0)[cc]{$R=L^\bot$}}
\put(79.00,20.00){\makebox(0,0)[cc]{$F$}}
\put(99.00,20.00){\makebox(0,0)[cc]{$B=F^\bot$}}
\end{picture}\medskip

d) On the left side below we depicted another orthospace and on the right
side the corresponding ortholattice.

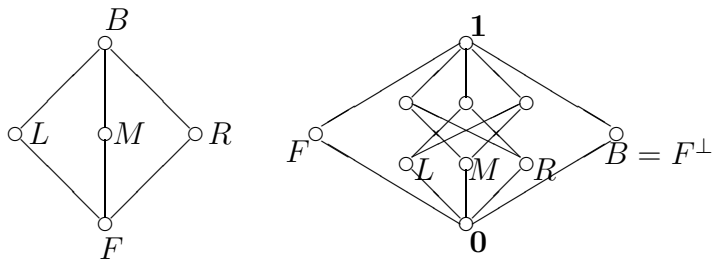
\unitlength=.8mm \special{em:linewidth 0.4pt} \linethickness{0.4pt} 
\begin{picture}(121.00,45.00)(-5,0)
\put(30.00,5.00){\circle{2.00}} \put(30.00,20.00){\circle{2.00}}
\put(15.00,20.00){\circle{2.00}} \put(45.00,20.00){\circle{2.00}}
\put(30.00,35.00){\circle{2.00}}
\put(31.00,1.00){\makebox(0,0)[cc]{$F$}}
\put(19.00,20.00){\makebox(0,0)[cc]{$L$}}
\put(34.00,20.00){\makebox(0,0)[cc]{$M$}}
\put(49.00,20.00){\makebox(0,0)[cc]{$R$}}
\put(32.00,39.00){\makebox(0,0)[cc]{$B$}}

 \put(16,21){\line(1,1){13}}  \put(16,19){\line(1,-1){13}}
\put(30,21){\line(0,1){13}}  \put(30,19){\line(0,-1){13}}
\put(44,21){\line(-1,1){13}}  \put(44,19){\line(-1,-1){13}}

\put(90.00,5.00){\circle{2.00}} \put(90.00,15.00){\circle{2.00}}
\put(100.00,15.00){\circle{2.00}} \put(80.00,15.00){\circle{2.00}}
\put(80.00,25.00){\circle{2.00}} \put(90.00,25.00){\circle{2.00}}
\put(100.00,25.00){\circle{2.00}} \put(90.00,35.00){\circle{2.00}}
\put(65.00,20.00){\circle{2.00}} \put(115.00,20.00){\circle{2.00}}

\put(89,5){\line(-5,3){23}} \put(91,5){\line(5,3){23}}
\put(89,35){\line(-5,-3){23}} \put(91,35){\line(5,-3){23}}
\put(89,6){\line(-1,1){8}} \put(91,6){\line(1,1){8}}
\put(90,6){\line(0,1){8}} \put(89,34){\line(-1,-1){8}}
\put(91,34){\line(1,-1){8}} \put(90,34){\line(0,-1){8}}
\put(81,16){\line(1,1){8}} \put(81,16){\line(2,1){18}}
\put(99,16){\line(-1,1){8}} \put(99,16){\line(-2,1){18}}
\put(89,16){\line(-1,1){8}} \put(91,16){\line(1,1){8}}

\put(62.00,17.00){\makebox(0,0)[cc]{$F$}}
\put(122.00,17.00){\makebox(0,0)[cc]{$B=F^\bot$}}
\put(83.00,14.00){\makebox(0,0)[cc]{$L$}}
\put(93.00,14.00){\makebox(0,0)[cc]{$M$}}
\put(103.00,14.00){\makebox(0,0)[cc]{$R$}}
\put(92.00,2.00){\makebox(0,0)[cc]{\bf 0}}
\put(92.00,38.00){\makebox(0,0)[cc]{\bf 1}}
\end{picture}\bigskip

We want to defend the thesis that ortholattices is a natural structure for
applying all the concepts that are used in the classical theory of
decision-making under uncertainty. As in the Boolean model we may speak of
the intersection ($\wedge$) and union ($\vee $), as well as of the
complementation (or as the negation, and understand it as
ortho-complementation). All the usual relations between these operations are
preserved with one exception: the law of distributivity is not satisfied in
the general case. But how often is it used? In the proofs of some theorems
and propositions, perhaps. But hardly in the formulation of the concepts.

A central point is that it is possible to speak about probabilities which
can be considered as a quantified saturation of the ortholattice skeleton.

\section{Non-classical probability}

We show here how the basic concepts of classical probability theory carry
over to ortholattices.

The theory of probability starts with the definition of a set $X$ of
elementary events. Thereafter it moves over to general events. In our
language events (or properties) are elements of an ortholattice $\mathcal{L}$%
. The next key concept is a "collection of mutually exclusive events". In
the classical model this is simply a partition of the set $X$, that is a
decomposition $X=A_{1}\amalg ...\amalg A_{n}$. In the general case the
notion of a collection of mutually exclusive events should be replaced by
the notion of an Orthogonal Decomposition of the Unit.\medskip

\textbf{Definition.} An \emph{Orthogonal Decomposition of the Unit} (ODU) in
an ortholattice $\mathcal{L}$ is a (finite) family of $\alpha =(a( i) ,\
i\in I(\alpha )$ ) of elements of $\mathcal{L}$ satisfying the following
condition: for any $i\in I(\alpha )$ 
\begin{equation*}
a( i) ^{\perp }=\bigvee _{j\neq i}a( j) .
\end{equation*}

The justification for this formulation is provided by that $a( i) \perp a(
j) \ $for\ $i\neq j$ and $\vee _{i}a( i) =\mathbf{1}$. The proof is obvious.

For instance, the single-element family \textbf{1 }is a (trivial) ODU. For
any $a\in \mathcal{L}$, the two-element family $( a,a^{\perp }) $ is an ODU.
We call this kind of family the \textit{question} about property $a$.

Intuitively, the family $\alpha $ is to be understood as a measurement (or a
source of information) with a set of possible outcomes $I( \alpha ) .$ If
such a measurement yields an outcome $i\in I( \alpha )$, we conclude that
our system is endowed with property $a( i) $ (or that the event $a(i)$
occurs). Assume that we can "prepare" our system in some state and
repeatedly measure the system (each time prepared in that same state) with
our measurement apparatus. The measurement outcomes can differ from one
trial to another. Imagine that we performed $n$ such measurements (for $n$
relatively large) and that outcome $i$ was obtained $n_{i}$ times. Then we
can assign each outcome $i$ a "probability" $p_{i}=n_{i}/n$. In fact we have
that $p_{i}\geq 0$ and $\sum p_{i}=1$. This leads us to\medskip

\textbf{Definition.} An \emph{evaluation} on an ortholattice $\mathcal{L}$
is a mapping $\nu: \mathcal{L} \to \mathbb{R}$. An evaluation $\nu$ is called

1)\emph{\ nonnegative} if $\nu ( a) \geq 0$ for any $a\in \mathcal{L};$

2) \emph{monotone} if $\nu ( a) \leq \nu ( b) $ $\ $when$\ a\leq b;\ $

3) \emph{normed} if\ $\nu ( \mathbf{1}) =1;$

4) \emph{additive} (or a \emph{measure}) if $\nu ( a\vee b) =\nu ( a) +\nu (
b) $ for orthogonal events $a$ and $b$. We write $a\oplus b$ instead of $%
a\vee b$ to emphasize that $a\perp b.$

5) \emph{probabilistic} (or a probability) if it is nonnegative and $%
\sum_{i}\nu (a(i)) =1$ for any ODU $( a( i),\ i\in I)$. \medskip

We make a few simple remarks on links between these concepts. From 4) or 5)
it follows easily that $\nu( \mathbf{0}) =0$; clearly then $%
2)\Longrightarrow 1)$. It is also clear that $5)\Longrightarrow 3)$, and 1),
3) and 4) together imply 5). In the classical (Boolean) case 5) implies 1) -
4), but that is not true in the general case. Indeed, let us consider
Example 3b, where (excluding the trivial events $\mathbf{\ 1\ }$and\textbf{\ 
}$\mathbf{0)}$ we have four events $r,\ l,\ r^{\perp },\ l^{\perp }$ and
where $r\leq l^{\perp }$ and $l\leq r^{\perp }$. To give a probability is
equivalent to give two numbers $\nu ( r) $ and $\nu (l)$ both between 0 and
1 but otherwise arbitrary. Such a probability is monotone if $\nu ( r) +\nu
( l) \leq 1$ and is additive if $\nu ( r) +\nu ( l) =1.$\medskip

There exists an important case when everything simplifies and approaches the
classical case. It is the case of orthomodular lattices. So are called the
lattices that satisfy the property of \textit{orthomodularity} (if $a\leq b$
then $b=a\vee ( b\wedge a^{\perp })$). It is clear that any Boolean lattice
is orthomodular and so are the lattices from Examples 2, 3c, and 3d. In
contrast, the lattice from Example 3b is not orthomodular. We assert that
for orthomodular lattices, property 5) implies 3) and 4).\medskip

\textbf{Lemma.} \emph{If $\mathcal{L}$\ is orthomodular ortholattice, then
any probability on $\mathcal{L}$ is additive and monotonic.}\medskip

Proof. Let $\nu$ be a probability on $\mathcal{L}$. We first establish
additivity. Suppose $a\perp b$ and pose $c=( a\oplus b) ^{\perp}$. Since $(
c,c^{\perp }) $ is an ODU, $\nu ( c) +\nu ( c^{\perp }) =1$.

We assert that $( a,b,c) $ is an ODU as well. To prove that we need to show
that $a^{\perp }=b\oplus c.$ Since $\ a,b$ and $c$ are pairwise orthogonal, $%
b\oplus c$ $\leq a^{\perp }.$\ By force of the property of orthomodularity
we have $a^{\perp }=( b\oplus c) \oplus ( a^{\perp }\wedge ( b\oplus c)
^{\perp }) $. But $a^{\perp }\wedge ( b\oplus c) ^{\perp }=( a\vee b\vee c)
^{\perp }=( a\oplus b) ^{\perp .}\wedge c^{\perp }=c\wedge c^{\perp }=%
\mathbf{0.}$\ Hence $a^{\perp }=b\oplus c$. Similarly $b^{\perp }=a\oplus c.$
The equality $c^{\perp }=a\oplus b$ is satisfied by definition. Thus, the
triplet $( a,b,c) $ is an ODU.

Therefore we have the equality $\nu ( a) +\nu ( b) +\nu ( c) =1$. Hence $\nu
( a\oplus b)=\nu(c^\perp) =1-\nu ( c) =\nu ( a) +\nu ( b)$, which yields the
additivity of $\nu$.

Monotonicity follows trivially from the formula $b=a\oplus ( b\wedge
a^{\perp })$, the additivity and the nonnegativity of the number $\nu (
b\wedge a^{\perp })$. QED\medskip

Thus, for the case of orthomodular lattices, a probability may also be
defined as a nonnegative normed measure.


\section{Qualitative Measures}

As it was already explained above we model uncertainty by an ortholattice of
properties or events. If we understand the elements of the lattice as
events, we may talk of smaller or larger probability for the realization of
these events. Further we focus on the "more (or less) likely than"
qualitative relation between events.\medskip

\textbf{Definition.} A \emph{qualitative measure} on an ortholattice $%
\mathcal{L}$ is a binary relation (of ``likelihood'') $\preceq $ on $%
\mathcal{L}$ satisfying the following two axioms:

QM1. \ \ $\preceq $\ is complete and transitive.

QM2. \ \ Let $a\preceq b$ and $a^{\prime }\preceq b^{\prime }$. Then $%
a\oplus a^{\prime }\preceq b\oplus b^{\prime }$ (recall that it means that $%
a\perp a^{\prime }$and$\ b\perp b^{\prime }$). The last inequality is strict
if at least one of the first inequalities is strict.\footnote{%
The special case of QM2 when $a^{\prime }=b^{\prime }$ is referred to in 
\cite{LS} as De Finetti axiom.} \medskip

A qualitative measure $\preceq $ \emph{is generated} by a (quantitative)
measure $\mu $ when $a\preceq b$ if and only if $\mu (a)\leq \mu (b)$. In
this section we are interested by the question as to when a qualitative
measure can be generated by a quantitative measure (or when there exists a
probabilistic sophistication). For simplicity we shall assume that the
ortholattice $\mathcal{L}$ is finite. But even in the classical context the
answer is generally negative (Kraft, Pratt, Seidenberg, 1959) Therefore in
order to obtain a positive answer we have to impose some additional
conditions which strengthen QM2. We shall here consider a condition
generalizing the classical \textquotedblleft cancellation
condition\textquotedblright . We prefer to call it \textquotedblleft
hyperacyclicity\textquotedblright . \medskip

\textbf{Definition.} A binary relation on $\mathcal{L}$ is said to be \emph{%
hyperacyclic} if the following condition holds:

Assume that we have a finite collection of pairs $( a_{i},b_{i})$ and that $%
a_{i}\preceq b_{i}$ for all $i$ and for some $i$ the inequality is strict.
Then $\sum \mu ( a_{i}) \neq \sum \mu ( b_{i}) $ $\ $for some measure $\mu $
on $\mathcal{L}$.\medskip

It is obvious that hyperacyclicity implies acyclicity as well as .

Clearly, if the qualitative relation $\preceq $ is generated by a measure $%
\mu $ then it is hyperacyclic. The main result of this section (and the
analog of Theorem 1 in \cite{LS}) asserts that for finite ortholattice the
reverse is true.\medskip

\textbf{Theorem 1.} \emph{Let $\preceq $ be a hyperacyclic qualitative
measure on a finite ortholattice $\mathcal{\ L}$. Then $\preceq $ is
generated by some measure on $\mathcal{L}$.}\medskip

A complete proof of Theorem 1 can be found in the Appendix. Here we confine
ourselves with describing the logic of the proof: We first embed the
ortholattice $\mathcal{L}$ into a vector space $V$ and identify linear
functionals on $V$ with measures on $\mathcal{L}$. With the qualitative
measure $\preceq $ we construct a subset $P\subset V$ and show that $0$ does
not belong to the convex hull of $P$. The separability theorem then
guarantees the existence of a linear functional on $V$ (that is of a measure
on $\mathcal{L}$) which is strictly positive on $P$. It is easy to show that
this measure generates the relation $\preceq $.\medskip

Clearly, if the relation $\preceq $ is monotonic (that is $a\preceq b$ for $%
a\leq b)$, then any measure $\mu $ generating $\preceq $ is also monotonic.
If, in addition, $\mathbf{0\prec 1}$ then $\mu ( \mathbf{1}) >0$; dividing
the measure $\mu $ by $\mu ( \mathbf{1}) $ we can assume that $\mu $ is a
normalized measure. Thus, the measure $\mu $ is a monotonic probability.

\section{Non-classical utility theory}

First of all we need to formulate a suitable generalization of the Savagian
concept of act. Roughly speaking an act is a bet on the result of some
measurement.\medskip

\textbf{Definition.} An \emph{act} is a pair $(\alpha ,f)$, where $\alpha
=(a(i),i\in I(\alpha))$ is some ODU (or a measurement), and $f:I(\alpha )
\to \mathbb{R}$ is a function.\medskip

We call the measurement $\alpha $ the \textit{basis} of our act.
Intuitively, if an outcome $i\in I( \alpha )$ is realized as a result of
measurement $\alpha $, then our agent receives $f(i)$ utils.

In such a way the set of acts with basis $\alpha $ can be identified with
the set (vector space, indeed) $F( \alpha ) =\mathbb{R}^{I( \mathbb{\alpha }%
) }$. The set of all acts $F$ is the disjoint union of $F( \alpha ) $ taken
over all ODU $\alpha .$

We are concerned with the comparison of acts with respect to their
attractiveness for our decision-maker. We start with a implicit formula for
such a comparison. Assume that the agent knows (more precisely, he thinks he
knows) the state of the system, that is he has in his mind a (subjective)
probability measure $\mu $ on the ortholattice $\mathcal{L}$. Then, for any
act $f$ on the basis $\alpha =(a(i),i\in I(\alpha ))$, he can compute the
following number (expected value of the act $f$) 
\begin{equation*}
CE_{\mu }(f)=\sum_{i}\mu (a(i))f(i).
\end{equation*}%
Using those numbers our agent can compare different acts.

We now shall (following Savage) go the other way around. We begin with a
relation $\preceq $ representing preferences over the set of all acts $F$,
thereafter we formulate axioms, impose conditions and arrive at the
conclusion that the preferences are explained by some probability measure $%
\mu $\ on $\mathcal{L}$.

More precisely, instead of a preference relation $\preceq $ on the set $F$
of acts, we at once assume the existence of a certainty equivalent $CE(f)$
for every act $f\in F$. (Of course that does simplify the task a little. But
this step is unrelated to the issue of classicality or non-classicality of
the "world"; it is only the assertion of the existence of a utility on the
set of acts. It would have been possible to obtain the existence of $CE$
from yet other axioms. We chose a more direct and shorter way).

Given that we shall only impose three requirements on $CE$. The first two
relate to acts defined on a fixed basis. Such acts are identified with
elements of the vector space $F(\alpha ) = \mathbb{R}^{\alpha }$.\medskip

\textbf{Monotonicity axiom.} The restriction of $CE$ on each $F(\alpha ) $
is a monotone functional.\medskip

\textbf{Linearity axiom.} For any measurement $\alpha $ the restriction of $%
CE$ on $F(\alpha)$ is a linear functional. \medskip

The third axiom links acts between different but in some sense comparable
basis. For this we need to be able to compare at least roughly two different
measurements. Consider two ODU $\alpha =(a(i),\ i\in I(\alpha))$ and $\beta
=(b( j) ,\ j\in I( \beta ) )$. We say the measurement $\alpha $ is \textit{%
finer} than $\beta $ if there exists a mapping $\varphi :I( \alpha ) \to I(
\beta )$ such that $a( i) \leq b( \varphi ( i) )$ for any $i\in I( \alpha )$%
. Simply stated it means that as we know a result $i$ of the first
measurement, we know the result of the second measurement without performing
it, it is $j=\varphi ( i) $. We note also that the transformation mapping $%
\varphi$ is uniquely defined. In fact assume that $\varphi ( i) $
simultaneously belongs to $b( j) $ and $b( k)$. Then $a( i) $ belongs to $b(
j) \wedge b( k)$. But since $b( j) $ and $b( k)$ are orthogonal $b( j)
\wedge b( k) =\mathbf{0}$, so $a( i) =0$. But this type of event do only
formally enter the decomposition of the unit and they can be neglected.

In any case any such mapping $\varphi :I(\alpha )\rightarrow I(\beta )$
defines a mapping 
\begin{equation*}
\varphi ^{\ast }:F(\beta )\rightarrow F(\alpha ).
\end{equation*}%
For a function $g$ on $I(\beta )$ the function $\varphi ^{\ast }(g)$ in a
point $i$ has the value $g(\varphi (i))$.

Intuitively, the payoffs from both functions (acts) $g$ and $f=\varphi
^{\ast }( g) $ are identical in all situations. Therefore our agent should
consider them as equivalent and assign them the same certainty equivalent.
This is the idea of the following axiom.\medskip

\textbf{Agreement axiom.} Suppose that a measurement $\alpha$ is finer than $%
\beta $ and $\varphi :I(\alpha) \to I(\beta )$ is the corresponding mapping.
Then $CE(g)=CE( \varphi ^*( g))$ for each $g\in F( \beta )$.\medskip

Take for instance $f$ to be the constant function in $I( \alpha ) $ with
value $1$. The agreement axiom says that the agent is indifferent between
two acts. The first is to receive one util without performing any
measurement. The second is to perform the measurement $\alpha $ and
(independently of the outcome) to receive a unit of utils.

The last requirement which cannot really be called an axiom says that the
utility of the trivial act with payoff 1 is equal to 1. That is $CE(1) =1.$%
\medskip

\textbf{Theorem 2.} \emph{Suppose that a certainty equivalent $CE$ satisfies
the monotonicity, linearity and agreement axioms. Then there exists a
probabilistic valuation $\mu$ on $\mathcal{L}$ such that $CE( f)
=\sum_{i}\mu ( a( i) ) f( i) $ for any act $f$ on the basis of measurement $%
\alpha =(a( i) ,\ i\in I( \alpha ))$. Moreover this valuation $\mu$ is
uniquely defined.}\medskip

\textit{Proof}. For $a\in \mathcal{L}$ we denote $1_{a}$ the bet on the
property $a$. It gives 1 util if we receive the answer YES on the question $%
(a,a^{\perp })$ and $0$ for NO. Let $\mu (a)=CE(1_{a})$. Since $1_{a}\geq 0$
we have $\mu (a)\geq 0$ for any $a\in \mathcal{L}$.

Let now $\alpha =(a( i) ,\ i\in I( \alpha ) )$ be an arbitrary ODU, and $%
f:I( \alpha ) \to \mathbb{R}$ be an act on the basis $\alpha $. We denote
with the symbol $1_i$ the act on the basis of $\alpha$ which yields $1$ on $%
i $ and 0 on $F( \alpha ) -\{ i\} $. By the agreement axiom we have that $%
CE(1_i) =\mu(a( i) ) $. Since $f=\sum_{i}f( i) 1_i$ we conclude that 
\begin{equation*}
CE( f) =\sum_{i}\mu ( a( i) ) f( i)
\end{equation*}%
In particular, if $f=1$ we obtain that $1=CE( 1) =\sum_{i}\mu ( a( i) ) $.
Therefore $\mu $ is a probabilistic valuation. QED\medskip

We do not assert that the valuation $\mu $ is monotone. In the next section
we substitute the agreement axiom with a stronger "dominance" axiom and we
obtain the monotonicity of $\mu$.

\section{The Dominance axiom}

Let $\alpha =(a( i), i\in I( \alpha ) )$ be a measurement (or an ODU). And
let $b\in \mathcal{L}$ be an event (or a property). We say that an outcome $%
i\in I(\alpha)$ is \emph{impossible under condition} $b$ (or in presence of
the property $b$), if $a( i) \perp b$. All other outcomes are in principle
possible, and we denote the set of possible outcomes as $I(\alpha|b)$.
Clearly 
\begin{equation*}
b\leq \bigvee_{i\in I(a\vert b) }a( i) =a( I( \alpha | b) ),
\end{equation*}
and $I(\alpha | b)$ is the smallest subset of $I( \alpha )$ with that
property. In fact if $b\leq a( J) $ then $a( J) ^{\perp }\leq b^{\perp }.$
But $a( J) ^{\perp }=a( I( \alpha ) -J) ,$ therefore for any $i,$ not
belonging to $J,$ we have $a( i) \leq b^{\perp },$ that is $a( i) \perp b$.

Consider for instance a situation when we have two measurements $\alpha =(a(
i) ,\ i\in I( \alpha ) )$ and $\beta =(b( j) ,\ j\in I( \beta ))$. Suppose
that the measurement $\alpha $ is finer than $\beta $ and $\varphi :I(
\alpha ) \to I(\beta ) $ is the corresponding mapping. Since 
\begin{equation*}
b( j) =a( \varphi ^{-1}( j) ) ,
\end{equation*}
it is easily seen that $I(\alpha \vert b( j) ) =\varphi^{-1}(j)$ and $%
I(\beta \vert a( i) ) =\{ \varphi ( i) \}$.

We go back to acts. Let $f:I( \alpha ) \to \mathbb{R}$ and $g:I( \beta ) \to 
\mathbb{R}$ be acts on the $\alpha $ and $\beta $ basis respectively. We say
the $g$ \textit{dominates} $f$ (and write $f\leq g)$ if for any $i\in I(
\alpha ) $ and any $j\in I(\beta \vert a( i) ) $ (that is $j$ is possible at
the event $a( i) )$ the inequality $f( i) \leq g( j) $ is true. Intuitively,
this means that the act $g$ always gives no less than the act $f$. With such
an interpretation it is natural to assume that our rational decision-maker
must assign to $g$ no less utility than to $f.$ We formulate this as\medskip

\textbf{Axiom of dominance.} If $f\leq g$ then $CE( f) \leq CE( g) $.\medskip

It is clear that the dominance implies monotonicity. We assert that the
dominance axiom also implies the axiom of agreement. In fact let $\beta $ be
a measurement coarser than $\alpha $ and $f=\varphi ^{\ast }( g) $ for some
act $g$ on the $\beta $ basis. From the description above it is clear that $%
f\leq g$ and $g\leq f$\ such that $CE( f) =CE( g)$.\medskip

\textbf{Theorem 3.} \emph{Assume that the axiom of linearity and dominance
are satisfied. Then $CE$ is an expected utility for some monotonic
probability measure $\mu$ on $\mathcal{L}$.}\medskip

\emph{Proof.} The first statement follows from earlier remarks and theorems.
Therefore we should prove the monotonicity of the measure $\mu $. Let $a\leq
b$. Consider two measurement-questions $\alpha =( a,a^{\perp }) $ and $\beta
=( b,b^{\perp }) .$ Let $f=1_{a,\text{ }}$that is a bet on event (property) $%
a$ $:$ the agent receives one util if measurement $\alpha $ reveals
(actualizes) property $a,$ and receives nothing in the opposite case. We
define $1_{b}\ $similarly on the $\beta $ basis. Clearly $1_{a}\leq 1_{b}.$
In fact if the first measurement reveals (actualizes) property $a,$ then $b$
is true for sure since $a\leq b.$\ Therefore $1_{b}$ gives the agent one
utils when $a$ occurs, and $\geq 0$ utils when $a^{\perp }$ occurs, which is
not worth less than 1$_{a}$. By force of the axiom of dominance $CE( \alpha
) \leq CE( \beta )$. The first term is equal to $\mu(a)$ and the second to $%
\mu(b)$. QED

\section*{Appendix}

Here we prove Theorem 1.\medskip

1. \emph{Construction of the vector space} $V$. Denote $\mathbb{R\otimes }%
\mathcal{L}$ the vector space generated by $\mathcal{L}$. It consists of
(finite) formal expressions of the form $\sum_{i}r_{i}a_{i}$, where $%
r_{i}\in \mathbb{R}$ and $a_{i}\in \mathcal{L}$. Denote $K$ the vector
subspace in $\mathbb{R\otimes }\mathcal{L}$ generated by expressions $%
a\oplus b-a-b$ (recall that $a\oplus b$ means that $a\oplus b=a\vee b$ and $%
a\perp b$.) Finally, $V=V(\mathcal{L})$ is the quotient space $\mathbb{%
R\otimes }\mathcal{L}$\ by the subspace $K$, $V=(\mathbb{R\otimes }\mathcal{L%
})/K$.

The ortholattice $\mathcal{L}$ naturally maps into $V$; the image $1\cdot a$
of an element $a\in \mathcal{L}$ we denote simply as $a$. Any linear
functional $l$ on $V$ restricted to $\mathcal{L}$ gives a valuation on $%
\mathcal{L}$. Since $l( a\oplus b-a-b) =l( a\oplus b) -l( a) -l( b) =0$, the
valuation $l$ is additive, that is a measure on the ortholattice $\mathcal{L}
$. Conversely, let $l$ be a measure on $\mathcal{L}$. We extend it by
linearity to $\mathbb{R\otimes }\mathcal{L}$ assuming $l( \sum r_{i}a_{i})
=\sum r_{i}l( a_{i})$. By force of additivity, $l$ yields $0$ for elements
of the form $a\oplus b-a-b$, that is $l$ vanishes on the subspace $K$.
Therefore $l$ factors through $V$ and is obtained from a linear functional
defined on $V$. We just proved\medskip

\textbf{Proposition 1.} \textit{The vector space of measures on $\mathcal{L}$
is identified with the space $V^{\ast }$ of linear functionals on }$V$%
.\medskip

Remark. The canonical mapping $\mathcal{L} \to V(\mathcal{L})$ can be
considered as the universal measure on the ortholattice $\mathcal{L}$. It is
injective if and only if the ortholattice $\mathcal{L}$ is
orthomodular.\medskip

2. \emph{Construction of the set of \textquotedblleft strictly
positive\textquotedblright\ $P$.} Let $\preceq $ be a binary relation on $%
\mathcal{L}$; as usual, $\prec $ denote the strict part of $\preceq $. By
definition, $P=P(\preceq )$ consists of (finite) expressions of the form $%
\sum_{i}(a_{i}-b_{i})$, where $b_{i}\preceq a_{i}$ for all $i$ and $%
b_{i}\prec a_{i}$ for some $i$. ($P$ is empty if the relation $\prec $ is
empty, that is if all elements in $\mathcal{L}$\ are equivalent relatively
to $\preceq $.) We note also that $P$ is stable with respect to the
addition.\medskip

3. Suppose now that a relation $\preceq $ is hyperacyclic. Note that the
hyperacyclicity of $\preceq $ means precisely that $0$ does not belongs to $%
P $.\medskip

\textbf{Proposition 2.} \emph{If the relation $\preceq$ is hyperacyclic then 
$0$ does not belong to the convex hull of $P$.}\medskip

\emph{Proof.} Assume that $0$ is a convex combination of elements of $P$, $%
0=\sum_i r_{i}p_{i}$, where $p_{i}\in P$, $r_{i}\geq 0$, and $%
\sum_{i}r_{i}=1 $. By Caratheodory's theorem we can assume that the $p_{i}$
are affinely independent (and therefore the coefficients $r_{i}$ are
uniquely defined). We assert that in this case the coefficients are \textit{%
rational} numbers.

It would be simplest to say that the set $P$ is defined over the field of
rational numbers. But it is not so easy to provide a precise meaning to it.
For that purpose we choose and fix some subset $L\subset \mathcal{L}$, such
that its image in $V$ is a basis of that vector space. We also choose a
subset $M$ of expressions of the form $a\oplus b-a-b,$ which constitute a
basis of the subspace $K$. The union of $L$ and $M$ is a basis of the vector
space $\mathbb{R\otimes }\mathcal{L}$. On the other side, $\mathcal{L}$ is a
basis of $\mathbb{R\otimes }\mathcal{L}$ as well. Since elements of $L\cup M$
are rational combinations of elements of the $\mathcal{L}$,basis elements of 
$\mathcal{L}$, in turn, can be rationally expressed in terms of $L\cup M$.
In particular, the images of elements of $\mathcal{L}_{{}}$ in $V$ are
rational combinations of elements of the $L$ basis. All the more, the
elements $p_{i}\in P$ can be rationally expressed in terms of $L$. It
follows (see, for example, Proposition 6 in \cite{Bour}, Chap. 2, \S\ 6)
that $0$ can be expressed rationally through $p_{i}$. Since the coefficients 
$r_{i}$ are defined uniquely, they are rational numbers.

Now the proof can be easily completed. We have an equality $0=\sum_i
r_{i}p_{i}$, where $p_{i}\in P$ and $r_{i}$ are rational numbers (not all
equal to zero). Multiplying with a suitable integer we may consider $r_{i}$
themselves as integers. Since $P$ is stable with respect to addition, we
obtain that $0\in P$, in contradiction with hyperacyclicity of the relation $%
\preceq $.\medskip

4. Together with Separation theorem of convex sets (see \cite{Roc}) the
results above imply existence of a (non-trivial) linear functional $\mu $ on 
$V$, non-negative on $P$. But we need strict positivity on $P$. To obtain it
we show that (in the case of a finite ortholattice $\mathcal{L}$) the convex
hull of $P$ is a polyhedron.

Let us introduce some notations. $A$ denotes the set of expression $a-b$,
where $a\succ b$. $B$ denotes the set of rays of the form $\mathbb{R}_{+}(
a-b) $, where $a\succeq b$. Finally, $Q$ is the convex hull of $A\cup B$ in $%
V$. By definition, $Q$ consists of elements of the form 
\begin{equation*}
q=\alpha _{1}( a_{1}-b_{1}) +...+\alpha _{n}( a_{n}-b_{n}) +\beta _{1}(
c_{1}-d_{1}) +...+\beta _{m}( c_{m}-d_{m}) ,\ \ \ \ \ \ \ \ ( \ast )
\end{equation*}
where $a_{i},b_{i},c_{j},d_{j}\in \mathcal{L\ }$ (more precisely, belong to
their image in $V$), $a_{i}\succ b_{i}$ for any $i$, $c_{j}\succeq d_{j}$
for any $j$, $\alpha _{i}, \ \beta _{i}$ are nonnegative, and $\sum_i \alpha
_{i}=1$.\medskip

\textbf{Proposition 3.} \emph{The convex hull of $\mathit{P}$ coincides with 
$Q$.} \medskip

\emph{Proof.} It is clear from the definitions that any element of $P$
belongs to $Q$. By the convexity of $Q$, the convex hull of $P$ is also
contained in $Q$.

It remains to show the converse, that any element $q$ of $Q$ belongs to the
convex hull of $P$. For that (appealing to the convexity of $co(P))$ we can
assume that $q$ has the form in $(\ast )$ with $n$ and $m$ equal to 1, that
is 
\begin{equation*}
q=(a-b)+\beta (c-d),
\end{equation*}%
where $a\succ b,\ c\succeq d\ $and $\beta \geq 0.$ If $\beta $ is an
integer, it is clear that $q\in P.$ In general case $\beta $ is a convex
combination of two nonnegative integers $\beta _{1}$ and $\beta _{2}$; then $%
q$ is the corresponding convex combination of two points $(a-b)+\beta
_{1}(c-d)$ and $(a-b)+\beta _{2}(c-d)$ both belonging to $P$.\medskip

\textbf{Corollary.} \emph{Assume that an ortholattice $\mathcal{L}$ is
finite. Then the convex hull of $P$ is a polyhedron.}\medskip

In fact, in this case the sets $A$ and $B$ are finite. Therefore (see \cite%
{Roc}, theorem 19.1) $Q$ is a polyhedra.

Thus, if 0 does not belong to the convex hull of $P$ (see Proposition 2)
then there exists a linear functional $\mu $ on $V$ which is strictly
positive on $P$. As we shall see, this immediately provides us with a proof
of Theorem 1.\medskip

5. \textit{Proof of Theorem 1}. The assertion in the theorem is trivially
true if all elements of $\mathcal{L}$ are equivalent to each other.
Therefore we can assume that there exists at least one pair $(a,b)$ such
that $a\succ b$. Let $\mu $ be a linear functional on $V$ (we may consider $%
\mu $ as a measure on the ortholattice $\mathcal{L}$) strictly positive on $%
P $. We assert that this measure generates the relation $\preceq .$

Let us suppose $c\succeq d$. Since for any integer positive number $n$ the
element $(a-b)+n(c-d)$ belongs to $P$, we have $\mu (a)-\mu (b)>n$ $(\mu
(d)-\mu (c))$ for any $n$. This implies $\mu (d)\leq \mu (c)$. Conversely,
let us suppose $\mu (c)\geq \mu (d)$ for some $c,d\in \mathcal{L}$. We have
to show that $c\succeq d$. If this is not the case then, by completeness of
the relation $\succeq $, we have $d\succ c$. But then $d-c$ belongs to $P$
and $\mu (d-c)=\mu (d)-\mu (c)>0$, which contradicts to our first
assumption. This completes the proof of Theorem 1.

\end{document}